\documentclass[conference]{IEEEtran}
\usepackage{cite}

\ifCLASSINFOpdf
\else
\fi
\usepackage{graphicx}

\graphicspath{{/home/ahsan/Documents/Ahsan_Research_Papers/PhD_papers/BDCloud_15/iiswc_paper/figures/}}
\DeclareGraphicsExtensions{.jpg}

\usepackage{array}

\usepackage{caption}
\ifCLASSOPTIONcompsoc
\usepackage[caption=false,font=normalsize,labelfon
t=sf,textfont=sf]{subfig}
\else
\usepackage[caption=false,font=footnotesize]{subfi
g}
\fi

\usepackage{stfloats}

\usepackage{multirow}
\usepackage[export]{adjustbox}

\begin{document}
%
\title{Performance Characterization of In-Memory Data Analytics on a Modern Cloud Server}


%
%
%
%


%
\author{

\IEEEauthorblockN{
Ahsan Javed Awan\IEEEauthorrefmark{1},  
Mats Brorsson\IEEEauthorrefmark{1}, 
Vladimir Vlassov\IEEEauthorrefmark{1} and 
Eduard Ayguade\IEEEauthorrefmark{2}
}

\IEEEauthorblockA{\IEEEauthorrefmark{1}KTH Royal Institute of Technology,\\
Software and Computer Systems Department(SCS),\\
\{ajawan, matsbror, vladv\}@kth.se}

\IEEEauthorblockA{\IEEEauthorrefmark{2}Technical University of Catalunya (UPC),\\
Computer Architecture Department,\\
eduard@ac.upc.edu}


}


\maketitle

\begin{abstract}

In last decade, data analytics have rapidly progressed from traditional disk-based processing to modern in-memory processing. 
However, little effort has been devoted at enhancing performance at micro-architecture level. 
This paper characterizes the performance of in-memory data analytics using Apache Spark framework. 
We use a single node NUMA machine and identify the bottlenecks hampering the scalability of workloads. We also  
quantify the inefficiencies at micro-architecture level for various data analysis workloads. 
Through empirical evaluation, we show that spark workloads do not scale linearly beyond twelve threads, due to work time inflation and thread level load imbalance. 
Further, at the micro-architecture level, we observe memory bound latency to be the major cause of work time inflation.

\end{abstract}


%
\IEEEpeerreviewmaketitle

\section{\textbf{Introduction}}

With a deluge in the volume and variety of data being collected at enormous rates, various enterprises, like Yahoo, Facebook and Google, are deploying clusters to run data analytics that extract valuable information from petabytes of data.
For this reason various frameworks have been developed to target applications in the domain of batch processing~\cite{hadoop}, graph processing~\cite{pregel} and stream processing~\cite{storm}. Clearly large clusters of commodity servers are the most cost-effective way to process exabytes but first, majority of analytic jobs do not process huge data sets~\cite{Scale_up_vs_Scale_out_for_Hadoop}. Second, machine learning algorithms are becoming increasingly common, which work on filtered datasets that can easily fit into memory of modern scale-up servers. Third, today's servers  can have substantial CPU, memory, and storage I/O resources. Therefore it is worthwhile to consider data analytics on modern scale-up servers.

In order to ensure effective utilization of scale-up servers, it is imperative to make a workload-driven study on the requirements that big data analytics put on processor and memory architectures. There have been several studies focusing on characterizing the behaviour of big data workloads and identifying the mismatch between the processor and the big data applications~\cite{Clearing_the_clouds,DCBench,BigDataBench,
understanding_in_memory_workloads, Characterising_and_subsetting_big_data_workloads, deep_dive_data_analytics,   MMU_Performance_Scale_out_workloads}. However, these studies lack in quantifying the impact of processor inefficiencies on the performance of in memory data analytics, which is impediment to propose novel hardware designs to increase the efficiency of modern servers for in-memory data analytics. To fill in this gap, we perform an extensive performance characterization of these workloads on a scale-up server using Spark framework. 

In summary, we make the following contributions:

\begin{itemize}

\item We perform an in-depth evaluation of Spark based data analysis workloads on a scale-up server. 
\item We discover that work time inflation (the additional CPU time spent by threads in a multi-threaded computation beyond the CPU time required to perform the same work in a sequential computation) and load imbalance on the threads are the scalability bottlenecks.
\item We quantify the impact of micro-architecture on the performance, and observe that DRAM latency is the major bottleneck.
\end{itemize}

\section{\textbf{Background}}

\subsection{\textbf{Spark}}

Spark is a cluster computing framework that uses Resilient Distributed Datasets (RDDs)~\cite{Spark}, which are immutable collections of objects spread across a cluster. Spark programming model is based on higher-order functions that execute user-defined functions in parallel. These higher-order functions are of two types: Transformations and Actions. Transformations are lazy operators that create new RDDs. Actions launch a computation on RDDs and generate an output. When a user runs an action on an RDD, Spark first builds a DAG of stages from the RDD lineage graph. Next, it splits the DAG into stages that contain pipelined transformations with narrow dependencies. Further, it divides each stage into tasks. A task is a combination of data and computation. Tasks are assigned to executor pool threads. Spark executes all tasks within a stage before moving on to the next stage. Table~\ref{spark_parameters} describe the parameters necessary to configure Spark properly in local mode on a scale-up server.

\begin{table}[h]
\renewcommand{\arraystretch}{1.3}
\caption{Spark Configuration Parameters}
\label{spark_parameters}
\centering
\begin{tabular}{m{3cm}|m{5cm}}
\hline
\textbf{Parameter}             & \textbf{Description}                                                                                     \\ \hline
spark.storage.memoryFraction   & fraction of Java heap to use for Spark's memory cache                                                    \\ \hline
spark.shuffle.compress         & whether to compress map output files                                                                     \\ \hline
spark.shuffle.consolidateFiles & whether to consolidates intermediate files created during a shuffle                                      \\ \hline
spark.broadcast.compress       & whether to compress broadcast variables before sending them                                              \\ \hline
spark.rdd.compress             & whether to compress serialized RDD partitions                                                            \\ \hline
spark.default.parallelism      & default number of tasks to use for shuffle operations (reduceByKey,groupByKey, etc) when not set by user \\ \hline
\end{tabular}
\end{table}

\subsection{\textbf{Top-Down Method for Hardware Performance Counters}}
Super-scalar processors can be conceptually divided into the "front-end" where instructions are fetched and decoded into constituent operations, and the "back-end" where the required computation is performed. A pipeline slot represents the hardware resources needed to process one micro-operation. The top-down method assumes that for each CPU core, there are four pipeline slots available per clock cycle. At issue point each pipeline slot is classified into one of four base categories: Front-end Bound, Back-end Bound, Bad Speculation and Retiring. If a micro-operation is issued in a given cycle, it would eventually either get retired or cancelled. Thus it can be attributed to either Retiring or Bad Speculation respectively. Pipeline slots that could not be filled with micro-operations due to problems in the front-end are attributed to Front-end Bound category whereas pipeline slot where no micro-operations are delivered due to a lack of required resources for accepting more micro-operations in the back-end of the pipeline are identified as Back-end Bound~\cite{Top_Down_Method_for_Counters}.

\section{\textbf{Methodology}}

\subsection{\textbf{Benchmarks}}

We select the benchmarks based on following criteria; (a) Workloads should cover a diverse set of Spark lazy transformations and actions, (b) Same transformations with different compute complexity functions should be included, (c) Workloads should be common among different Big Data Benchmark suites available in the literature.(d) Workloads have been used in the experimental evaluation of Map-Reduce frameworks for Shared-Memory Systems.

Table~\ref{Benchmarks} shows the list of benchmarks along with transformations and actions involved.  Most of the workloads have been used in popular data analysis workload suites such as BigDataBench~\cite{BigDataBench}, DCBench~\cite{DCBench}, HiBench~\cite{HiBench} and Cloudsuite~\cite{Clearing_the_clouds}. Phoenix++~\cite{Phoenix++}, Phoenix rebirth~\cite{Phoenix_Rebirth} and Java MapReduce~\cite{Java_MapReduce} tests the performance of devised shared-memory frameworks based on Word Count, Grep and K-Means. We use Spark version of the selected benchmarks from BigDataBench and employ Big Data Generator Suite (BDGS), an open source tool, to generate synthetic datasets for every benchmark based on raw data sets~\cite{BDGS}. We work with smaller datasets deliberately to fully exploit the potential of in-memory data processing.

\begin{itemize}
\item \textbf{Word Count (Wc)} counts the number of occurrences of each word in a text file. The input is unstructured Wikipedia Entries.

\item \textbf{Grep (Gp)} searches for the keyword "The" in a text file and filters out the lines with matching strings to the output file.  It works on unstructured Wikipedia Entries.
\item \textbf{Sort (So)} ranks records by their key. Its input is a set of samples. Each sample is represented as a numerical d-dimensional vector.

\item \textbf{Naive Bayes (Nb)} uses semi-structured Amazon Movie Reviews data-sets for sentiment classification. We use only the classification part of the benchmark in our experiments.

\item \textbf{K-Means (Km)} clusters data points into a predefined number of clusters. We run the benchmark for 4 iterations with 8 desired clusters. Its input is structured records, each represented as a numerical d-dimensional vector.

\end{itemize}

\begin{table}[h!]
\renewcommand{\arraystretch}{1.3}
\caption{Benchmarks}
\label{Benchmarks}
\centering
\begin{tabular}{p{2cm}p{1.5cm}|p{1.9cm}|p{1.9cm}}
\hline
\multicolumn{2}{l|}{\textbf{Benchmarks}} & \textbf{Transformations} & \textbf{Actions} \\ \hline
\multicolumn{1}{l|}{Micro-benchmarks} & Word count & map & saveAsTextFile \\
\multicolumn{1}{l|}{} &  & reduceByKey &  \\ \cline{2-4} 
\multicolumn{1}{l|}{} & Grep & filter & saveAsTextFile \\ \cline{2-4} 
\multicolumn{1}{l|}{} & Sort & map & saveAsTextFile \\
\multicolumn{1}{l|}{} &  & sortByKey &  \\ \hline
\multicolumn{1}{l|}{Classification} & Naive Bayes & map & collect \\
\multicolumn{1}{l|}{} &  &  & saveAsTextFile \\ \hline
\multicolumn{1}{l|}{Clustering} & K-Means & map & takeSample \\
\multicolumn{1}{l|}{} &  & mapPartitions & collectAsMap \\
\multicolumn{1}{l|}{} &  & reduceByKey & collect \\
\multicolumn{1}{l|}{} &  & filter &  \\ \hline
\end{tabular}
\end{table}

\subsection{\textbf{System Configuration}}

Table~\ref{hardware} shows details about our test machine. Hyper-Threading and Turbo-boost are disabled through BIOS because it is difficult to interpret the micro-architectural data with these features enabled~\cite{HT_disabled}. With Hyper-Threading and Turbo-boost disabled, there are 24 cores in the system operating at the frequency of 2.7 GHz.

\begin{table}[h]
\renewcommand{\arraystretch}{1.3}
\caption{System}
\label{hardware}
\centering
\begin{tabular}{l|l|p{4.7cm}}
\hline
\textbf{Component} & \multicolumn{2}{l}{\textbf{Details}} \\ \hline
Processor & \multicolumn{2}{l}{Intel Xeon E5-2697 V2, Ivy Bridge micro-architecture} \\ \hline
\multirow{6}{*}{} & Cores & 12 @ 2.7 GHz (Turbo up 3.5 GHz) \\ \cline{2-3} 
 & Threads & 2 per Core \\ \cline{2-3} 
 & Sockets & 2 \\ \cline{2-3} 
 & L1 Cache & 32 KB for Instruction and 32 KB for Data per Core \\ \cline{2-3} 
 & L2 Cache & 32 KB per core \\ \cline{2-3} 
 & L3 Cache (LLC) & 30 MB per Socket \\ \hline
Memory & \multicolumn{2}{l}{2 x 32 GB, 4 DDR3 channels, Max BW 60 GB/s} \\ \hline
OS & \multicolumn{2}{l}{Linux Kernel Version 2.6.32} \\ \hline
JVM & \multicolumn{2}{l}{Oracle Hotspot JDK 7u71} \\ \hline
\end{tabular}
\end{table}

Table~\ref{parameters} also lists the parameters of JVM and Spark. For our experiments, we use HotSpot JDK version 7u71 configured in server mode (64 bit). The heap size is chosen to avoid getting "Out of memory" errors while running the benchmarks. The open file limit in Linux is increased to avoid getting "Too many files open in the system" error. The young generation space is tuned for every benchmark to minimize the time spent both on young generation and old generation garbage collection, which in turn reduces the execution time of the workload. The size of young generation space and the values of Spark internal parameters after tuning are available in Table~\ref{parameters}.
   
\begin{table}[!ht]
\renewcommand{\arraystretch}{1.3}
\caption{JVM and Spark Parameters for Different Workloads.}
\label{parameters}
\centering
\begin{tabular}{p{0.45cm}|l|ccccc}
\hline
\multicolumn{2}{c|}{\textbf{Parameters}} & \multicolumn{1}{c|}{\textbf{Wc}} & \multicolumn{1}{c|}{\textbf{Gp}} & \multicolumn{1}{c|}{\textbf{So}} & \multicolumn{1}{c|}{\textbf{Km}} & \textbf{Nb} \\ \hline
\multirow{5}{*}{JVM} & Heap Size (GB) & \multicolumn{5}{c}{50} \\ \cline{2-7} 
 & Young Generation Space (GB) & \multicolumn{1}{c|}{45} & \multicolumn{1}{c|}{25} & \multicolumn{1}{c|}{45} & \multicolumn{1}{c|}{15} & 45 \\ \cline{2-7} 
 & MaxPermSize (MB) & \multicolumn{5}{c}{512} \\ \cline{2-7} 
 & Old Generation Garbage Collector & \multicolumn{5}{c}{ConcMarkSweepGC} \\ \cline{2-7} 
 & Young Generation Garbage Collector & \multicolumn{5}{c}{ParNewGC} \\ \hline
\multirow{7}{*}{Spark} & spark.storage.memoryFraction & \multicolumn{1}{c|}{0.2} & \multicolumn{1}{c|}{0.2} & \multicolumn{1}{c|}{0.2} & \multicolumn{1}{c|}{0.6} & 0.2 \\ \cline{2-7} 
 & spark.shuffle.consolidateFiles & \multicolumn{5}{c}{true} \\ \cline{2-7} 
 & spark.shuffle.compress & \multicolumn{5}{c}{true} \\ \cline{2-7} 
 & spark.shuffle.spill & \multicolumn{5}{c}{true} \\ \cline{2-7} 
 & spark.shuffle.spill.compress & \multicolumn{5}{c}{true} \\ \cline{2-7} 
 & spark.rdd.compress & \multicolumn{5}{c}{true} \\ \cline{2-7} 
 & spark.broadcast.compress & \multicolumn{5}{c}{true} \\ \hline
\end{tabular}
\end{table}

\subsection{\textbf{Measurement Tools and Techniques}}

We use jconsole to measure time spent in garbage collection. We rely on the log files generated by Spark to calculate the execution time of the benchmarks. We use Intel Vtune~\cite{Vtune} to perform concurrency analysis and general micro-architecture exploration. For scalability study, each benchmark is run 10 times within a single JVM invocation and the median values of last 5 iterations are reported. For concurrency analysis, each benchmark is run 3 times within a single JVM invocation and Vtune measurements are recorded for the last iteration. This experiment is repeated 3 times and the best case in terms of execution time of the application is chosen. The same measurement technique is also applied in general architectural exploration, however the difference is best case is chosen on basis of IPC. Additionally, executor pool threads are bound to the cores before collecting hardware performance counter values. Although this measurement method is not the most optimal for Java experiments as suggested by Georges et al~\cite{Java_performance_evaluation}, we believe, it is enough for Big Data applications. We use a top-down analysis method proposed by Yasin~\cite{Top_Down_Method_for_Counters} to identify the micro-architectural inefficiencies.   

\subsection{\textbf{Metrics}}

The definition of metrics used in this paper, are taken from Intel Vtune online help~\cite{Vtune}.
\begin{itemize}
\item \textbf{CPU Time:} is time during which the CPU is actively executing your application on all cores.
\item \textbf{Wait Time:} occurs when software threads are waiting on I/O or due to synchronization.
\item \textbf{Spin Time:} is wait time during which the CPU is busy. This often occurs when a synchronization API causes the CPU to poll while the software thread is waiting.
\item \textbf{Core Bound:} shows how core non-memory issues limit the performance when you run out of out-of-order execution resources or are saturating certain execution units.
\item \textbf{Memory Bound:} measures a fraction of cycles where pipeline could be stalled due to demand load or store instructions.
\item \textbf{DRAM Bound:} shows how often CPU was stalled on the main memory.
\item \textbf{L1 Bound:} shows how often machine was stalled without missing the L1 data cache. 
\item \textbf{L2 Bound:} shows how often machine was stalled on L2 cache. 
\item \textbf{L3 Bound:} shows how often CPU was stalled on L3 cache, or contended with a sibling Core.
\item \textbf{Store Bound:} This metric shows how often CPU was stalled on store operations.
\item \textbf{Front-End Bandwidth:} represents a fraction of slots during which CPU was stalled due to front-end bandwidth issues.
\item \textbf{Front-End Latency:} represents a fraction of slots during which CPU was stalled due to front-end latency issues.
\end{itemize}
 
\section{\textbf{Scalability Analysis}}

In this section, we evaluate the scalability of benchmarks. Speed-up is calculated as $T_1$/$T_n$, where $T_1$ is the execution time with a single executor pool thread, and $T_n$ is the execution time using $n$ threads in the executor pool. 

\subsection{\textbf{Application Level}}

\begin{figure}[h!]
\includegraphics[scale=0.38,center]{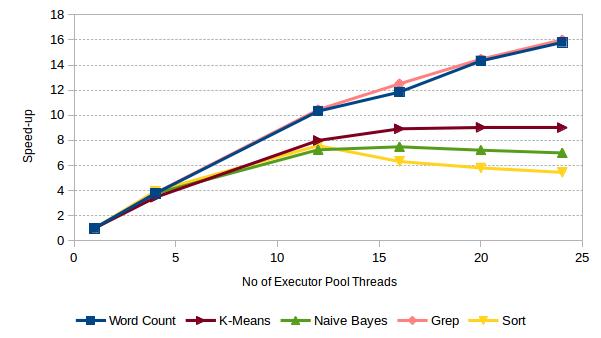}
\caption{Scalability of Spark  Workloads in Scale up Configuration}
\label{speedup_applicaiton}
\end{figure}

Figure~\ref{speedup_applicaiton} shows the speed-up of workloads for increasing number of executor pool threads. All workloads scale perfectly up to 4 threads. From 4 to 12 threads, they show linear speed-up. Beyond 12 threads, Word Count and Grep scale linearly but the speed-up for Sort, K-Means and Naive Bayes tend to saturate.

\subsection{\textbf{Stage Level}}

Next we drill down to stage level and observe how different stages scale with the number of executor pool threads. We only study those stages whose execution time contributes to 5\% of total execution time of workload, e.g Naive Bayes has 2 stages but only the stage Nb\_1 contributes significantly to the total execution time. Grep has only a filter stage, Word Count has a map stage (Wc\_1) and a reduce stage (Wc\_2). In Sort, So\_0 and So\_3 are map stages, So\_1 is SortByKey stage and sorted data is written to local file in So\_2 stage. In K-Means, map stages are Km\_0, Km\_18, Km\_20, Km\_22, Km\_22, Km\_24 and Km\_26. Km\_1 and Km\_27 are takeSample and sum stages. Stages Km\_3, Km\_4, Km\_6, Km\_7, Km\_9, Km\_10, Km\_12, Km\_13, Km\_15 and Km\_16 perform mapPartitionswithIndex transformation. Stages up to Km\_18 belong to initialization phase whereas the remaining ones belong to the iteration phase of K-Means.

\begin{figure*}[!t]
\centering
\subfloat[Word Count, Naive Bayes, Grep and Sort]{\includegraphics[scale=0.38]{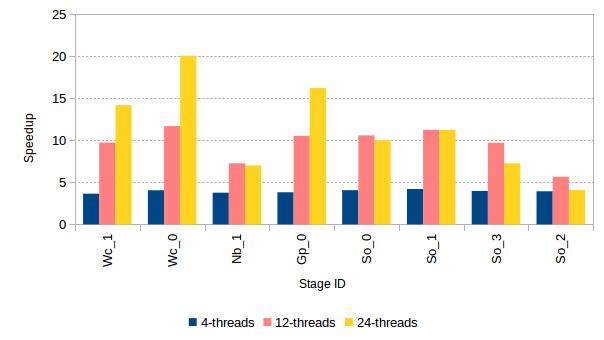}
\label{stages_wc_so_gp_nb}}
\subfloat[K-Means]{\includegraphics[scale=0.38]{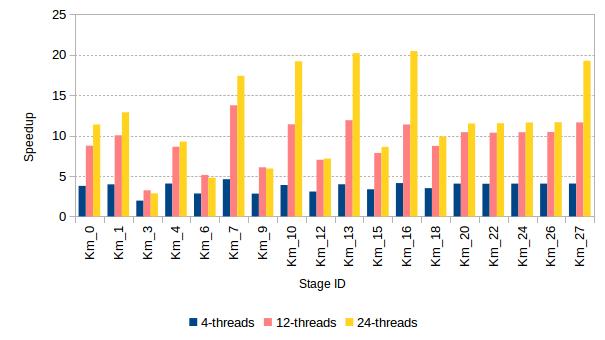}
\label{stages_km}}
\caption{Performance at Stage Level}
\label{performance_task_level}
\end{figure*}

At 4-threads case (see Figure ~\ref{stages_wc_so_gp_nb}), all stages of a workload exhibit ideal scalability but in 12 and 24-threads, the scalability characteristics vary among the stages, e.g. Wc\_0 shows better speed-up than Wc\_1 in 24-threads case. The scalability of Sort is worst among all applications in 24-threads case because of So\_2 stage that does not scale beyond 4 threads. In K-Means (see Figure~\ref{stages_km}), stages where mapPartitionswithIndex transformations are performed show better scalability than map stages both in the initialization and iteration phases. The scalability of map transformations vary, e.g in 24-threads case, map stage in Word Count  has better scalability than that in Sort, Naive Bayes and K-Means.This can be attributed to the complexity of user defined functions in map transformations.

\subsection{\textbf{Tasks Level}}

\begin{figure*}[!t]
\centering
\subfloat[Word Count (Wc\_1)]{\includegraphics[scale=0.38]{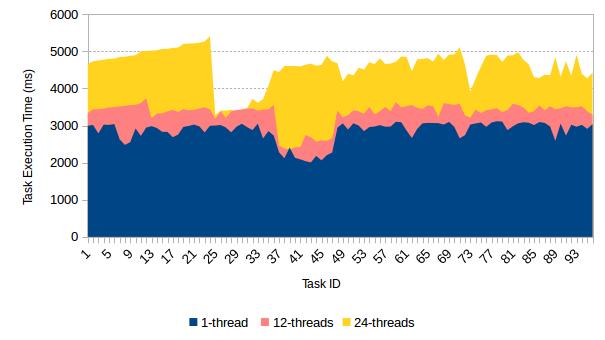}
\label{tasks_execution_wc_1}}
\subfloat[Kmeans (Km\_0)]{\includegraphics[scale=0.38]{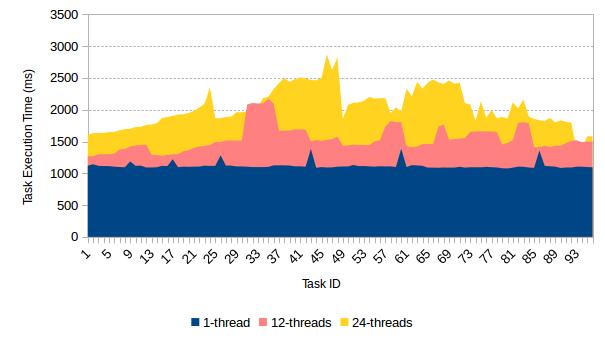}
\label{tasks_execution_Km_0}}
\caption{Performance at Task Level}
\label{performance_task_level}
\end{figure*}

Figure~\ref{tasks_execution_wc_1} and~\ref{tasks_execution_Km_0} show the execution time of tasks in Wc\_1 and Km\_0 stage respectively. Note that the size of task set does not change with increase in threads in the executor pool because it depends on the size of input data set. The data set is split into chunks of 32 MB by default. The figures show that execution time of tasks increases with increase in threads in the executor pool. To quantify the increase, we calculate area under the curves (AUC) using trapezoidal approximation. Table~\ref{auc} presents percentage increase in AUC for various workloads in multi-threaded cases over 1-thread case. For Wc\_1, there is 17\% and 61\% increase in AUC 12-threads and 24-threads case over 1-thread case. For So\_3, there is 24\% and 68\% increase where as for Km\_0, the increase is 38\% and 83\%

\begin{table}[h!]
\renewcommand{\arraystretch}{1.3}
\caption{Percentage increase in AUC compared to 1-thread}
\label{auc}
\centering
\begin{tabular}{p{2cm}|p{1.75cm}p{1.75cm}}
\hline
\textbf{Stage} & \textbf{12-threads} & \textbf{24-threads} \\ \hline
Wc\_1 & 17.03      & 61.50      \\
So\_3 & 24.58      & 68.50      \\
Km\_0 & 38.02      & 83.20     \\ \hline
\end{tabular}
\end{table}

\section{\textbf{Scalability Limiters}}

\subsection{\textbf{CPU Utilization}}

\begin{figure}[h!]
\centering
\includegraphics[scale=0.38]{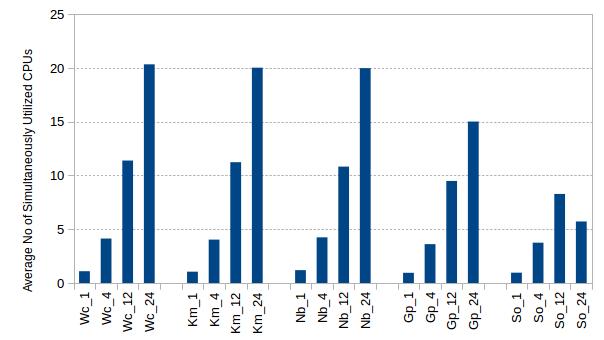}
\caption{CPU Utilization of Benchmarks}
\label{cpu_usage}
\end{figure}

Figure~\ref{cpu_usage} shows the average number of CPU's used during the execution time of benchmarks for different number of threads in the executor pool. By comparing this data with speed-up numbers in Figure~\ref{speedup_applicaiton}, we see a strong correlation between the two for 4-threads case and 12-threads case. At 4-threads case, 4 cores are fully utilized in all benchmarks, At 12-threads case, Word Count, K-Means and Naive Bayes utilize 12 cores, whereas Grep and Sort utilize 10 and 8 cores respectively. At 24-threads case, none of the benchmarks utilize more than 20 cores. This utilization further drop to 16 for Grep and 6 for Sort. The performance numbers scale accordingly  for these two benchmarks but for Word Count, K-Means and Naive Bayes, the performance is not scaling along with CPU utilization. We try to answer why such behaviour exists on these programs in subsequent sections

\subsection{\textbf{Load Imbalance on Threads}}

\begin{figure*}[!t]
\centering
\subfloat[K-Means]{\includegraphics[scale=0.38]{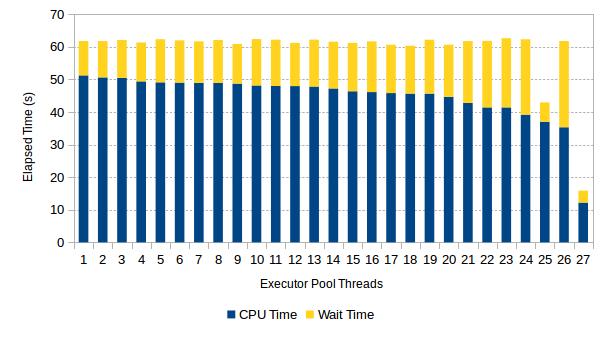}
\label{li_km_24}}
\hfil
\subfloat[Variation from Mean CPU Time for Different No of Executor Pool Threads]{\includegraphics[scale=0.38]{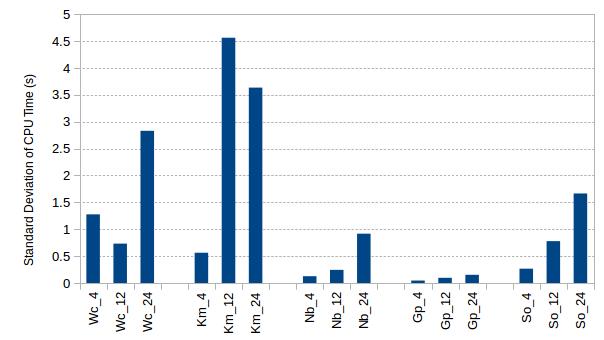}
\label{standard_deviation}}
\caption{Load Imbalance in Spark Benchmarks}
\label{load_imbalance}
\end{figure*}

Load imbalance means that one or a few executor pools threads need
(substantially) more CPU time than other threads, which limits the achievable speed-up, as the threads with less CPU time will have more wait time and if the CPU time across the threads is balanced, over-all execution time will decrease. Figure~\ref{li_km_24} breaks down elapsed time of each executor pool thread in K-Means in to CPU time and wait time for 24-threads case. The worker threads are shown in descending order of CPU time. The figure shows load imbalance. To quantify load imbalance, we compute the standard deviation of CPU time and show for 4, 12 and 24-threads case for all benchmarks in  Figure~\ref{standard_deviation}. The problem of load imbalance gets severe at higher number of threads. The major causes of load imbalance are; a non uniform division of the work among the threads,resource sharing, cache coherency or synchronization effects through barriers~\cite{speedup_stacks}. 

\subsection{\textbf{Work Time Inflation}}

\begin{figure*}[!t]
\centering
\subfloat[K-Means]{\includegraphics[scale=0.38]{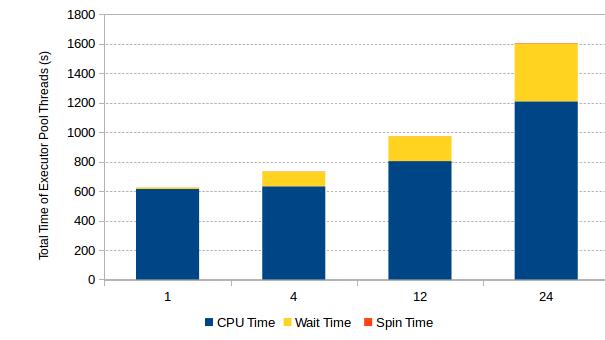}
\label{total_time_km}}
\hfill
\subfloat[Elapsed Time Breakdown]{\includegraphics[scale=0.38]{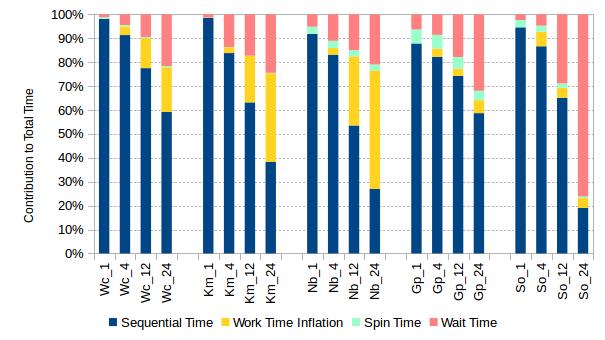}
\label{elapsed_time_breakdown}}
\caption{Work Time Inflation in Spark Benchmarks}
\label{performance_thread_level}
\end{figure*}

In this section, we drill down at threads level and analyse the behaviour of only executor pool threads because they contribute to 95\% of total CPU time during the entire run of benchmarks. By filtering out executor pool threads in the concurrency analysis of Intel Vtune, we compute the total CPU time, spin time and wait time of worker threads and the numbers are shown in Figure~\ref{total_time_km} for K-Means at 1, 4, 12 and 24-threads case. The CPU time in 1-thread case is termed as sequential time, the additional CPU time spent by threads in a multi-threaded computation beyond the CPU time required to perform the same work in a sequential computation is termed as work time inflation as suggested by Oliver et-al~\cite{worktime_inflation}. 

Figure~\ref{elapsed_time_breakdown} shows the percentage contribution of sequential time, work time inflation, spin time and wait time towards the elapsed time of applications. The spinning overhead is not significant since it contribution is less than  around 5\% across all workloads in both sequential and multi-threaded cases. The contribution of wait time tends to increase with increase in threads in the executor pool. The percentage fractions are increased by, 20\% in Word Count and K-Means, 15\% in Naive Bayes, 25\% in Grep and 70\% in Sort. Word Count, K-Means and Naive Bayes see increase in fraction of work time inflation with increase in threads in the executor pool. At 24-threads, the contribution of work time inflation is 20\%, 36\% and 51\% in Word count, K-Means and Naive Bayes respectively. For Grep and Sort, this overhead is between 5-6\% at 24-threads case. 

By comparing the data in Figure~\ref{performance_thread_level} with performance data in Figure~\ref{speedup_applicaiton}, we see that Grep does not scale because of wait time overhead. Sort has the worst scalability because of significant contribution of wait time. In Word Count, there is equal contribution of work time inflation and wait time overhead where as K-Means and Naive Bayes are mostly dominant by work time inflation. Moreover the work time inflation overhead also correlates with speed-up numbers, i.e. Word Count having less work time inflation scales better than K-Means and Naive Bayes having largest contribution of work time inflation scales poorer than K-Means. In the next section, we try to find out the micro-architectural reasons that result in work time inflation.

\subsection{\textbf{Micro-architecture}}

\begin{figure*}[!t]
\centering
\subfloat[IPC]{\includegraphics[scale=0.38]{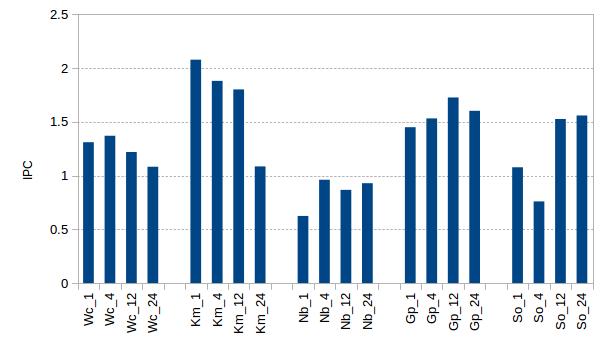}
\label{ipc}}
\hfil
\subfloat[Top Level]{\includegraphics[scale=0.38]{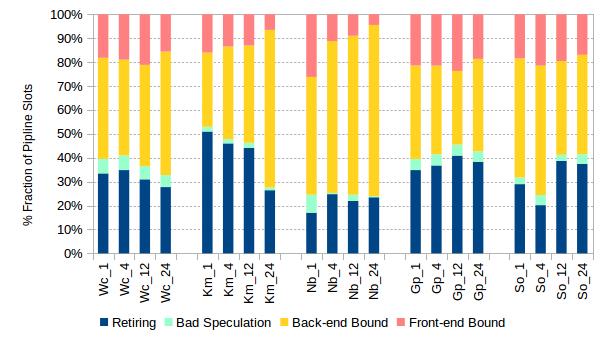}
\label{top_level}}
\hfil
\subfloat[Backend Level]{\includegraphics[scale=0.38]{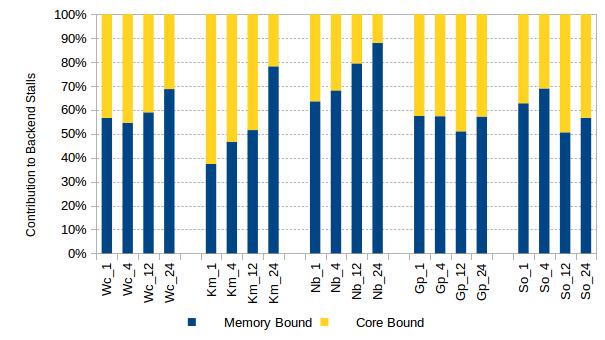}
\label{backend_level}}
\hfil
\subfloat[Memory Level]{\includegraphics[scale=0.38]{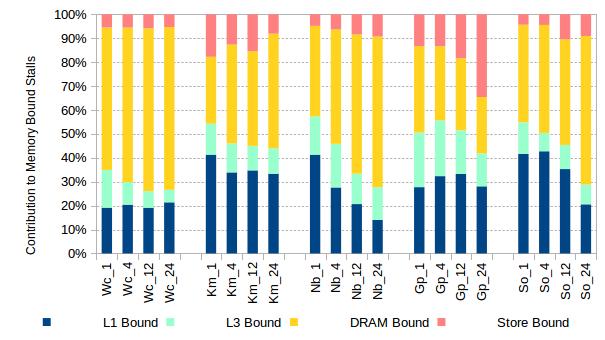}
\label{memory_level}}
\hfil
\subfloat[Core Level]{\includegraphics[scale=0.38]{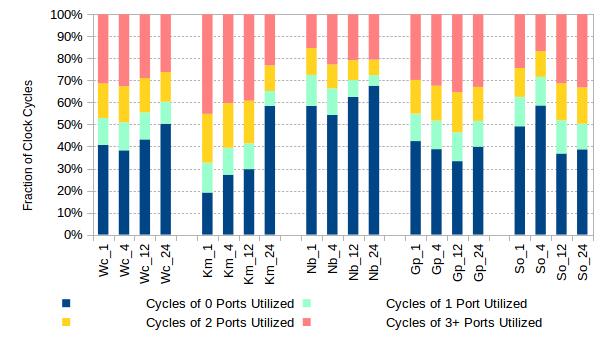}
\label{core_level}}
\hfil
\subfloat[Frontend Level]{\includegraphics[scale=0.38]{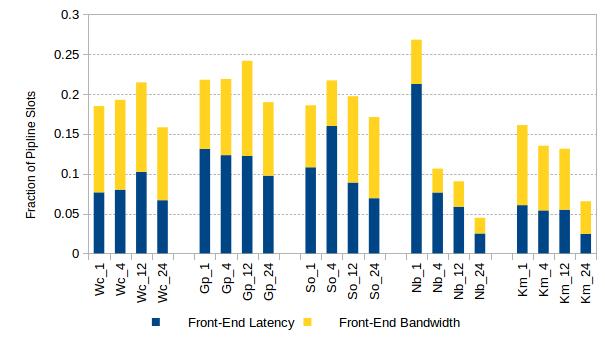}
\label{frontend_level}}
\caption{Top-Down Analysis Breakdown for Benchmarks with Different No of Executor Pool Threads}
\label{top_down_analysis}
\end{figure*}

\paragraph{\textbf{Top Level}} Figure~\ref{top_level} shows the breakdown of pipeline slots for the benchmarks running with different number of executor pool threads. On average across the workloads; Retiring category increases from  33.4\% in 1-thread case to 35.7\% in 12-threads case (Note how well it correlates to IPC) and decreases to 31\% in 24-threads case,  Bad Speculation decreases from 4.7\% 1-threads case to 3.1\% in 24-threads case, Front-end bound decreases from 20.4\% in 1-thread case to 12.6\% in 24-threads case and Back-end bound increases from 42.9\% in 1-thread case to 54.3\% 12-threads case. This implies that workloads do not scale because of issues at the Back-end. The contribution of Back-end bound increases with increase in number of worker threads in workloads suffering with work time inflation and in 24-threads case, it correlates with speed-up, i.e. the higher the Back-end bound is, the lower the speed-up is.

\paragraph{\textbf{Backend Level}} Figure~\ref{backend_level} shows the contribution of memory bound stalls and core bound stalls. On average across the workloads; the fraction of memory bound stalls increases from 55.6\% in 1-thread case to 72.2\% in 24-threads. It also shows that workloads exhibiting larger memory bound stalls results in higher work time inflation.

\paragraph{\textbf{Memory Level}} Next we drill down into Memory level in Figure~\ref{memory_level}. The Memory level breakdown suggests that on average across the workloads, fraction of L1 bound stalls decrease from 34\% to 23\%, fraction of L3 bound stalls decrease from 16\% to 10\%, fraction of Store bound stalls increase from 9\% to 11\% and the fraction of DRAM bound stalls increase 42\% to 56\%, when comparing the 1-thread and 24-threads cases. The increase in fraction of DRAM bound stalls correlate to work time inflation,  30\% increase in DRAM bound stalls yields higher work time inflation Naive Bayes that K-Means for 24-threads case where increase in contribution of DRAM bound stalls is 20\%. Word Count with only 10\% increase in DRAM bound stalls shows exhibit lower amount of work time inflation than K-Means.

\paragraph{\textbf{Execution Core Level}} Figure~\ref{core_level} shows the utilization of execution resources for benchmarks at multiple no of executor pool threads. On average across the workloads, the fraction of clock cycles during which no port is utilized (execution resources were idle) increases from 42.3\% to 50.7\%, fraction of cycles during which 1, 2 and 3 + ports are used decrease from 13.2\% to 8.9\%, 15.7\% to 12.8\% and 29.3\% to 27.1\% respectively, while comparing 1 and 24-threads case.  
 
\paragraph{\textbf{Frontend Level}} Figure~\ref{frontend_level} shows the fraction of pipeline slots during which CPU was stalled due to front-end latency and front-end bandwidth issues. At higher number of threads, front- end stalls are equally divided among latency and bandwidth issues. On average across the workloads; front-end latency bound stalls decrease from 11.8\% in 1-thread case to 5.7\% in 24-threads case where as front-end bandwidth bound stalls decrease from 8.6\% to 6.9\%.  

\subsection{\textbf{Memory Bandwidth Saturation}}

\begin{figure}[h!]
\centering
\includegraphics[scale=0.38]{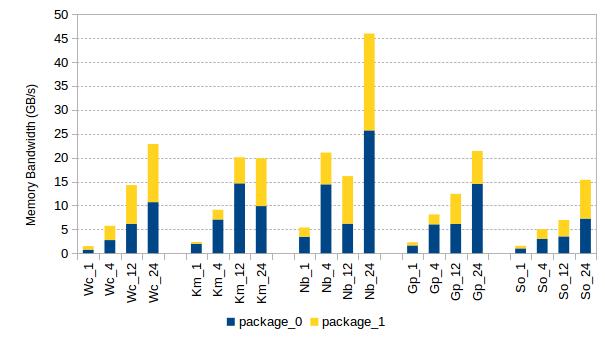}
\caption{Memory Bandwidth Consumption of Benchmarks}
\label{memory_bandwidth}
\end{figure}

Figure~\ref{memory_bandwidth} shows the amount of data read and written to each of the two DRAM packages via the processor's integrated memory controller. The bandwidth (Gigabytes/sec) to package\_1 shows an increasing trend with increase in threads in the executor pool. The same trend can be seen for total memory bandwidth in most of the workloads. We also see an imbalance between memory traffic to two DRAM packages. Off-chip bandwidth requirements of  Naive Bayes are higher than rest of the workloads but the peak memory bandwidth of all the workloads are with in the platform capability of 60 GB/s, hence we conclude that memory bandwidth is not hampering the scalability of in-memory data analysis workloads.

\section{\textbf{Related Work}}
Oliver et al.~\cite{worktime_inflation} have shown that task parallel applications can exhibit poor performance due to work time inflation. We see similar phenomena in Spark based workloads. Ousterhout et al.~\cite{Making_sense} have developed blocked time analysis to quantify performance bottlenecks in the Spark framework and found out that CPU (and not I/O) is often the bottleneck. Our thread level analysis of executor pool threads also reveal that CPU time (and not wait time) is the dominant performance bottleneck in Spark based workloads.

Ferdman et al.~\cite{Clearing_the_clouds} show that scale-out workloads suffer from high instruction-cache miss rates. Large LLC does not improve performance and off-chip bandwidth requirements of scale-out workloads are low. Zheng et al.~\cite{OS_behaviour_scale_out_workloads} infer that stalls due to kernel instruction execution greatly influence the front end efficiency. However, data analysis workloads have higher IPC than scale-out workloads~\cite{DCBench}. They also suffer from notable from end stalls but L2 and L3 caches are effective for them. Wang et al.~\cite{BigDataBench} conclude the same about L3 caches and L1 I Cache miss rates despite using larger data sets. Deep dive analysis~\cite{deep_dive_data_analytics} reveal that big data analysis workload is bound on memory latency but the conclusion can not be generalised. None of the above mentioned works consider frameworks that enable in-memory computing of data analysis workloads.

Jiang et al.~\cite{understanding_in_memory_workloads} observe that memory access characteristics of the Spark and Hadoop workloads differ. At the micro-architecture level, they have roughly same behaviour and point current micro-architecture works for Spark workloads. Contrary to that, Jia et al.~\cite{Characterising_and_subsetting_big_data_workloads} conclude that Software stacks have significant impact on the micro-architecture behaviour of big data workloads. However both studies lack in quantifying the impact of micro-architectural inefficiencies on the performance. We extend the literature by identifying the bottlenecks in the memory subsystem.  

\section{\textbf{Conclusion}}

We evaluated the performance of Spark based data analytic workloads on a modern scale-up server at application, stage, task and thread level. While performing experiments on a 24 core machine, we found that that most of the applications exhibit sub-linear speed-up, stages with map transformations do not scale, and execution time of tasks in these stages increases significantly. The CPU utilization for several workloads is around 80\% but the performance does not scale along with CPU utilization. Work time inflation and load imbalance on the threads are the scalability bottlenecks. We also quantified the impact of micro-architecture on the performance. Results show that issues in front end of the processor account for up to 20\% of stalls in the pipeline slots, where as issues in the back end account for up to 72\% of stalls in the pipeline slots. The applications do not saturate the available memory bandwidth and memory bound latency is the cause of work time inflation. We will explore pre-fetching mechanisms to hide the DRAM access latency in data analysis workloads, since Dimitrov et al.~\cite{Memory_system_characterization_big_data_workloads} show potential for aggressively pre-fetching large sections of the dataset onto a faster tier of memory subsystem.

\bibliographystyle{IEEEtran}
\bibliography{IEEEabrv,references}

\end{document}